\documentstyle[12pt,psfig]{article}
\def\lf{\leaders\hbox to 1em{\hss.\hss}\hfill}
\def\21{$SU(2) \ot U(1)$}
\def\321{$SU(3) \ot SU(2) \ot U(1)$}
\def\ne{\hbox{$\nu_e$ }}

\def\nt{\hbox{$\nu_\tau$ }}

\def\eq#1{{eq. (\ref{#1})}}
\def\Eq#1{{Eq. (\ref{#1})}}

\def\VEV#1{\left\langle #1\right\rangle}
\let\vev\VEV

\def\lsim{\raise0.3ex\hbox{$\;<$\kern-0.75em\raise-1.1ex
\hbox{$\sim\;$}}}
\def\gsim{\raise0.3ex\hbox{$\;>$\kern-0.75em\raise-1.1ex
\hbox{$\sim\;$}}}

\def\beq{\begin{equation}}
\def\eeq{\end{equation}}
\def\bef{\begin{figure}}
\def\eef{\end{figure}}
\def\bet{\begin{table}}
\def\eet{\end{table}}
\def\bea{\begin{eqnarray}}
\def\eea{\end{eqnarray}}
\def\ba{\begin{array}}
\def\ea{\end{array}}
\def\bi{\begin{itemize}}
\def\ei{\end{itemize}}
\def\ben{\begin{enumerate}}
\def\een{\end{enumerate}}
\def\ra{\rightarrow}
\def\ot{\otimes}

\def\apj#1#2#3{   { Astrophys. J. }{\bf #1}, #2 (19#3)}

\def\np#1#2#3{    { Nucl. Phys. }{\bf #1}, #2 (19#3)}
\def\pl#1#2#3{    { Phys. Lett. }{\bf #1}, #2 (19#3)}
\def\pr#1#2#3{    { Phys. Rev. }{\bf #1}, #2 (19#3)}
\def\prep#1#2#3{  { Phys. Rep. }{\bf #1}, #2 (19#3)}
\def\prl#1#2#3{   { Phys. Rev. Lett. }{\bf #1}, #2 (19#3)}

\def\Zp#1#2#3{    { Z. Physik }{\bf #1}, #2 (19#3)}

\def\n.c.#1#2#3{  { Nuovo Cim. }{\bf #1}, #2 (19#3)}
\def\r.n.c.#1#2#3{{ Riv. del Nuovo Cim. }{\bf #1}, #2 (19#3)}
\def\sjnp#1#2#3{  { Sov. J. Nucl. Phys. }{\bf #1}, #2 (19#3)}

\def\mpl#1#2#3{   { Mod. Phys. Lett. }{\bf #1}, #2 (19#3)}

\renewcommand{\thefootnote}{\fnsymbol{footnote}}
\def\dfrac#1#2{{\displaystyle\frac{#1}{#2}}}

\begin{document}
\thispagestyle{empty}
\begin{titlepage}
\begin{center}
\rightline{hep-ph/9606445}
\hfill FTUV/96-33\\
\hfill IFIC/96-39\\
\hfill \today \\
\vskip 0.2cm
{\Large \bf  Supernova Bounds on Supersymmetric $R$-parity Violating 
Interactions \\ 
}
\vskip 0.2cm
{H.  Nunokawa}$^1$ 
\footnote{
E-mail: nunokawa@flamenco.ific.uv.es},
A. Rossi$^1$
\footnote{
E-mail: rossi@evalvx.ific.uv.es, rossi@ferrara.infn.it},
and 
J. W. F. Valle$^1$
\footnote{
E-mail: valle@flamenco.ific.uv.es}\\
{\sl $^1$Instituto de F\'{\i}sica Corpuscular - C.S.I.C.\\
Departament de F\'{\i}sica Te\`orica, Universitat de Val\`encia\\
46100 Burjassot, Val\`encia, SPAIN\\
URL http://neutrinos.uv.es}\\
\vskip .2cm
{\bf Abstract}
\end{center}
\begin{quotation}

We re-examine resonant massless-neutrino conversions in a 
dense medium induced by flavour changing neutral current (FCNC) 
interactions.
We show how the observed $\bar\nu_e$ energy spectra from SN1987a 
and the supernova $r$-process nucleosynthesis  provide 
constraints on supersymmetric models with $R$ parity violation, 
which are much more stringent than those obtained from the laboratory.
We also suggest that resonant massless-neutrino conversions may 
play a positive role in supernova shock reheating.
Finally, we examine the constraints on explicit $R$-parity-violating  
FCNCs in the presence of non-zero neutrino masses in the eV range, 
as indicated by present hot dark matter observations.

\end{quotation}
\end{titlepage}

\renewcommand{\thefootnote}{\arabic{footnote}}
\setcounter{footnote}{0}
\section{Introduction}

Under suitable circumstances neutrinos can 
oscillate in the presence of matter \cite{W} or
undergo resonant conversions  \cite{valle} even when 
they are strictly massless. In some models even
unmixed neutrinos can resonantly convert in 
matter \cite{ER,GMP}. Massless-neutrino resonant conversions 
are distinct from the usual MSW conversions \cite{W,MS} in that
they are independent of neutrino energy and affect
simultaneously neutrinos as well as anti-neutrinos. 
For this reason this mechanism is expected to play an 
important role in supernova physics \cite{valle,AS,NQRV}.
Moreover, in some of these models with flavour changing 
neutrino neutral current (FCNC) interactions with matter 
constituents it has been suggested that, for a certain 
range of the corresponding parameters, they may account for
the observed deficit of solar neutrinos \cite{ER,GMP,BPW}.
 
The required ingredients can naturally emerge in the context 
of various models beyond the standard model \cite{beyond}. 
In particular, in this paper we consider this type of 
interactions mediated by the scalar partners of quarks 
and leptons in  supersymmetric extensions of the standard 
model with explicitly broken $R$ parity \cite{SW,rparity}. 

The presence of $R$ parity breaking interactions
induce resonant neutrino conversions of the type
${\nu}_e \leftrightarrow {\nu}_\alpha$ 
as well as  $\bar{\nu}_e \leftrightarrow \bar{\nu}_\alpha$. 
Such conversions have important implications for 
the supernova $r$-process nucleosynthesis \cite{QFMMW}
as well as the observed $\bar{\nu}_e$ energy 
spectra  from SN1987a \cite{old,SSB,JNR}. 

In a recent work \cite{NQRV}, we have investigated
the constraints on massless neutrino resonant conversions 
that follow from supernova considerations. In the present 
paper we apply the same considerations in order to constrain
models with explicit $R$ parity violating supersymmetric 
interactions which can effectively induce resonant 
conversions even when neutrino masses are neglected.
We also suggest that resonant massless-neutrino 
conversion may play a positive role in supernova shock reheating.
In addition, we generalize this approach in order to
include the possibility of non-zero neutrino masses.
These are typically expected to arise in these models
and could help to explain present observations.
We derive the corresponding constraints on flavour 
changing neutral current couplings generated 
by explicit $R$ parity violating interactions.

The paper is structured as follows. 
In Sect. 2  we briefly present the form of the FCNC and flavour 
diagonal neutral current (FDNC) interactions emerging from the 
$R$ parity violating terms and the new effective neutrino 
evolution Hamiltonian in matter. In particular we consider 
two possible scenarios:
\begin{enumerate}
\item  massless and unmixed neutrinos 
($\delta m^2 =0, \sin 2\theta=0$) with 
FCNC as well as {\it non standard} FDNC  
interactions of neutrinos with matter;
\item massive neutrinos ($\delta m^2 \neq 0$) assuming negligible 
mixing in vacuum  ($\sin 2\theta=0$), but with FCNC interactions.  
\end{enumerate}
Sect. 3 is devoted to a discussion of resonant massless-neutrino 
conversions for supernova neutrino detection and $r$-process
nucleosynthesis. We show how the observed $\bar\nu_e$ energy 
spectra from SN1987a and the supernova $r$-process nucleosynthesis 
place important restrictions on the parameters of $R$ parity 
violating models. In sect. 4 we discuss the second scenario 
above and derive the corresponding  restrictions. In sect. 5
we briefly suggest resonant massless-neutrino 
conversion as a way to power supernova shock reheating.
Finally, we summarize our results and conclude in Sect. 6.
 
\section{The MSW effect with FCNC interactions}

 $R$ parity is a quantum number which is +1 for all standard 
particles and -1 for the super partners. It is directly related 
to the baryon ($B$) and lepton ($L$) number as 
$R= (-1) ^{3B +L + 2S}$, where $S$ is the particle spin. 
In the Minimal Supersymmetric Standard Model (MSSM) \cite{mssm}  
the $R$ discrete symmetry is imposed to enforce the 
$L$ and $B$ number conservation 
and no tree-level flavour changing interactions exist. 
However no fundamental principle precludes the possibility 
to violate these symmetries \cite{SW,rparity}. 
Within the particle content of the MSSM 
$R$ parity can be broken explicitly by renormalizable 
(and hence {\sl a priori} unsuppressed) operators. 
The following extra $L$ violating couplings 
in the superpotential are directly relevant for neutrino
propagation through matter:
\bea
\label{lepton}
\lambda_{ijk} L_i L_j E^c_k \, \\
\lambda'_{ijk} L_i Q_j D^c_k 
\label{lq}
\eea
where $L, Q, E^c$ and  $D^c$ are (chiral) superfields which 
contain the usual lepton 
and quark $SU(2)$  doublets and singlets, respectively, and $i,j,k$ are 
generation indices. 
In the next we focus only on the second term \eq{lq}
because  the first is much more constrained by experimental data.
Note that the simultaneous presence of the 
 $\lambda'' U^c U^c D^c$ and $\lambda' L Q D^c$-type couplings 
is very strongly 
constrained ($\lambda',\lambda'' \leq 10^{-10}$) 
from non-observation of proton decay \cite{BGH,VS}.
However, the constraints on $\lambda'$ (see below) are rather weak
in the absence of the $B$ violating $\lambda''_{ijk}U^c_i U^c_jD^c_k$
term. We will adopt this choice throughout this paper.

 The couplings in \eq{lq} 
at low energy ($< 100$GeV) give rise to the following 
four-fermion effective Lagrangian for neutrinos interactions with $d$-quark
\footnote{
For simplicity 
we omit in the $\lambda'$-type Yukawa couplings
the terms $\lambda'_{i1k} (\bar{\nu_{i L}})^c d_{1 L} (\tilde{d}_{k R})^*$ 
($i,k=1,2,3$). However, the coupling constants 
$\lambda'_{i1k}$ are much more 
constrained than  $\lambda'_{ik1}$ \cite{BGH}.}:
\beq
\label{effec}
L_{eff}  =  - 2\sqrt{2} G_F \sum_{\alpha,\beta} 
\xi_{\alpha\beta} \: \bar{\nu}_{L\alpha} \gamma^{\mu} \nu_{L\beta} \:
\bar{d}_{R}\gamma^{\mu}{d}_{R}\:\:\:\alpha,\beta = e,\mu, \tau \, , 
\eeq
where the parameters $\xi_{\alpha\beta}$ represent the strength of the 
effective interactions normalized to the Fermi constant $G_F$.  
For our purpose 
we consider explicitly the following {\it non standard} FDNC couplings:
\bea
\xi_{ee} &= &\sum_j \frac{|\lambda'_{1j1}|^2}{4 \sqrt{2} G_F m^2_
{\tilde{q}_{jL}}} 
 \, ,\\
\xi_{\mu\mu} &= &\sum_j \frac{|\lambda'_{2j1}|^2}
{4 \sqrt{2} G_F m^2_{\tilde{q}_{j L}}} \, ,\\
\xi_{\tau\tau}& = &\sum_j \frac{|\lambda'_{3j1}|^2}
{4 \sqrt{2} G_F m^2_{\tilde{q}_{j L}} }\, ,\,\,\, j = 1,2,3\, , 
\eea
 and the FCNC ones:
\bea
\label{fcnc}
\xi_{e\mu}& = &\sum_j \frac{\lambda'_{1j1}\lambda'_{2j1}}
{4 \sqrt{2} G_F m^2_{\tilde{q}_{j L}}} \, ,\\
\xi_{e\tau}& = &\sum_j \frac{\lambda'_{1j1}\lambda'_{3j1}}
{4 \sqrt{2} G_F m^2_{\tilde{q}_{j L}}}\, ,
\,\,\,\, j = 1,2,3\, , 
\eea
where $m_{\tilde{q}_{j L}}$ 
are the masses of the exchanged squarks and 
$j = 1,2,3$ denotes  $\tilde{d}_L, \tilde{s}_L, \tilde{b}_L$, respectively. 
These effective neutral current interactions contribute to the neutrino 
scattering off $d$ quarks in matter, providing new 
flavour conserving as well as flavour changing terms 
for the matter potentials of neutrinos. 

The phenomenological implications of the $R$ parity violating 
couplings have been extensively studied and constraints on the 
coupling constants $\lambda'$ from low-energy processes (charged 
current universality, $e-\mu-\tau$ universality, 
$\nu_\mu-e$ scattering, atomic parity violation)
has been obtained \cite{BGH}. Recently, new bounds have been 
derived from LEP electroweak observables to constrain $\lambda'_{i3k}$
 (for all $i,k$) and from $D$- decays to constrain $\lambda'_{12k}$ 
and $\lambda'_{22k}$ as well as from $\tau$ decays to restrict 
$\lambda'_{31k}$ (for all $k$) (see \cite{GB} and refs. therein). 
In summary, the most stringent bounds on   
the coupling constants entering our study 
are the following 
\footnote{In ref. \cite{KP} stringent bounds, 
$\lambda'_{113}\lambda'_{131}\leq 1.1\times 10^{-7}$, 
$\lambda'_{112}\lambda'_{121}\leq 3.2\times 10^{-5}$, 
$\lambda'^2_{111} \leq 6.4\times 10^{-5}$, 
are obtained from the non-observation of $0\nu\beta\beta$ decay
for squark masses of 100 GeV. However, these limits suffer from
some theoretical uncertainties on nuclear matrix elements.}
(at 1 $\sigma$ level): 
\bea
\label{boun}
\lambda'_{12k} &\leq 0.29\,,& \: \: \lambda'_{13k} \leq 0.26, \\\nonumber
\lambda'_{22k} &\leq 0.18\,,&\,   \: \: \lambda'_{23k} \leq 0.44,\\ \nonumber
\lambda'_{33k} &\leq 0.26\,,&\,   \: \:\hfil\\ \nonumber
\lambda'_{i1k} &\leq 0.05\,,&\,\,\, (i=1,2,\: k= 1,2,3) \, 
\eea
 normalized to a 100 GeV {\sl reference} squark mass. 

The most general Schroedinger neutrino evolution equation 
in matter takes the form 
\begin{equation}
{i{d \over dr}\left(\matrix{
\nu_e \cr\ \nu_x \cr }\right)=
  \left(\matrix{
 {H}_{e}
& {H}_{ex} \cr
 {H}_{ex} 
& {H}_{x} \cr}
\right)
\left(\matrix{
\nu_e \cr\ \nu_x \cr}\right) }\,, \,\,\,x =\mu (\tau)
\label{evolution1}
\end{equation}
The entries of the Hamiltonian reads as
\beq
\label{hamil}
H_e =  V_e - \frac{\delta m^2}{2E} \cos 2\theta \, , \,\,\, \,\, \,\,
H_{x} = V_x \, ,  \,\,\, \, \,\,   
H_{ex} =  V_{ex} +\frac{\delta m^2}{4E} \sin 2\theta 
\eeq
where $E$ is the neutrino energy,  
$\delta m^2$ is the mass squared difference, $\theta$ is 
the neutrino mixing angle in vacuum and $V_e, V_x$ and 
$V_{ex}$ are the effective matter potentials as given by
\bea
\label{poten}
V_e &=& \frac{\sqrt{2} G_F \rho}{m_p} 
\Bigl[\frac{3Y_e - 1}{2}  +  \xi_{ee} (2-Y_e)\Bigr]\, ,\\
V_x &= &\frac{\sqrt{2} G_F \rho}{m_p} 
\Bigl[\frac{Y_e-1}{2} +  \xi_{xx} (2-Y_e)\Bigr]\, ,\\
V_{ex}& = &\frac{\sqrt{2} G_F \rho}{m_p} \xi_{ex} (2-Y_e) \, .
\eea
Here $m_p$ is the nucleon mass, $\rho$ is the matter density, 
$Y_e$ is the electron number per nucleon and charge neutrality is 
assumed
\footnote{Here the $d$ quark number density $N_d$ in the 
medium is understood  to be expressed as $N_e + 2N_n$.}.
For the corresponding anti-neutrino states the sign of matter potentials 
is opposite. 

Let us note that the matter potential induced by the {\it non standard}  
FDNC interactions plays the role of an extra effective mass,  
whereas those induced by the FCNC couplings play the role of a new 
{\it mixing} term. 
As a result, in principle even for strictly massless neutrinos 
($\delta m^2=0$) and vanishing $\theta$, these new matter potentials 
make the resonant neutrino conversion in medium possible
\cite{W,ER,GMP}.

Let us now turn to the application 
of the above to the neutrino conversions in a supernova. 
Let us discuss  separately the cases 
of $\delta m^2 =0$ and $\delta m^2 \neq 0$.

\section{Massless neutrino resonant conversion in supernovae}

We now turn to the application of the previous 
formalism to resonant neutrino conversion in supernovae.

By equating the diagonal terms in the Hamiltonian matrix of 
\eq{hamil} one can infer, for the case of massless neutrinos, 
that the resonance condition is given by
\beq
\label{rc}
\xi'\equiv \xi_{xx} -\xi_{ee} = \frac{Y_e}{2-Y_e}
\eeq
which is clearly energy independent. 
Here we should note that a positive value of $\xi'$ is necessary for
the above equation to hold. 
It is important to note that 
the same resonance condition holds also for the anti-neutrino system 
$\bar{\nu}_e \leftrightarrow \bar{\nu}_x$. As a result, {\sl both} 
neutrinos {\sl and} anti-neutrinos can simultaneously undergo 
resonant conversions as discussed in ref. \cite{valle}. 
As a result, this can affect in an important way 
supernova neutrino emission.

The mixing angle  $\theta_m$ and the
neutrino oscillation length $L_m$ in matter are given by
\bea
%
%
L_m & = & 
 \frac{\pi\sin 2\theta_m}{V_{ex}}\, , \\
  \label{length}
\label{angle}
\tan 2\theta_m &
= & \frac{ 2 \xi_{ex} (2-Y_e)}{Y_e -\xi' (2-Y_e)}\, , 
\eea
respectively.

In our subsequent discussion, we will employ the simple Landau-Zener 
approximation \cite{Landau,HPD} to estimate the conversion 
probability after the neutrinos cross the resonance. Under this 
approximation, the probability for $\nu_e\leftrightarrow\nu_x$
and $\bar\nu_e\leftrightarrow\bar\nu_x$ conversions is given by
\begin{eqnarray}
\label{LZ}
  P & = & 1 - 
\exp\Biggl(-\dfrac{\pi^2}{2}\dfrac{\delta r}
{L_{m}^{\rm res}} \Biggr) \nonumber \\
          & \approx &  1 - \exp\left[
-5 \times 
10^4 \times\left(\dfrac{\rho_{\rm res}}{10^{12} {\mbox{g/cm}^3}}\right)
            \Biggl(\dfrac{h}{\mbox{cm}} \Biggr)
\dfrac{\xi^2_{ex}}{\xi'}
                 \right], \nonumber \\
 \delta r & = & 4 h \dfrac{\xi_{ex}}{\xi'}, \,\,\,\,\,\,\,\,
h \equiv \left| \frac{\mbox {d}\ln Y_e}{\mbox{d}r}\right|^{-1}_{\rm res}, 
\end{eqnarray}
where $L_{m}^{\rm res}$ is the neutrino oscillation length
at resonance and $\rho_{\rm res}$ is the corresponding matter density. 

Let us briefly review the supernova process we are going to consider. 
A few seconds after the bounce, the electron number density $Y_e$ 
is very low  just above the neutrinosphere, $Y_e \sim 10^{-2}$, 
while at large radii it saturates to an asymptotic value 
$\sim 0.4$ (see Sect. 4.1 in (\cite{NQRV}). 
This implies, from \eq{rc}, that the resonance condition 
requires lepton universality to be at least violated at the 1\% 
level, $\xi' \gsim 10^{-2}$ 
which is not in contradiction with present bounds outlined in \eq{boun}. 
To keep the discussion simple and more conservative, we consider,  
for each flavour conversion ($\nu_e\ra\nu_\mu$ or $\nu_e\ra\nu_\tau$),  
only the contribution due to the exchange of one left-handed $\tilde{q}$ 
at a time in the corresponding 
effective couplings $\xi_{ee}, \xi_{xx}, \xi_{ex}$.  

After the bounce of the core, all neutrinos, emitted from the 
neutrinosphere, have approximately equal luminosities 
but rather different energy spectra. Correspondingly, the 
average neutrino energies satisfy the following hierarchy:
\begin{equation}
\label{hierarchy}
\langle E_{\nu_e} \rangle <\langle E_{\bar\nu_e}\rangle <
\langle E_{\nu_{\tau(\mu)}}\rangle 
\approx\langle E_{\bar\nu_{\tau(\mu)}}\rangle.
\end{equation}
Typically, the average supernova neutrino energies are: 
\begin{equation}
\label{average}
\langle E_{\nu_e}\rangle \approx 11\ \mbox{MeV},\ \langle E_{\bar\nu_e}\rangle
\approx 16\ \mbox{MeV},\ \langle E_{\nu_{\tau(\mu)}}\rangle \approx \langle
E_{\bar\nu_{\tau(\mu)}}\rangle\approx 25\ \mbox{MeV}.
\end{equation} 
As a result, a considerable conversion $\bar{\nu}_e \leftrightarrow 
\bar{\nu}_{\mu,\tau}$ leads to a permutation of the neutrino energy
spectra which would provide a high energy tail in the anti-neutrino
energy spectrum from the supernova SN1987a \cite{KA,IMB}. 
Comparison with the SN1987A observations leads to an upper bound  
for the transition probability $P$ close to 0.35 \cite{SSB}. 
Following the same reasoning,  we will constrain the effective FCNC 
couplings that can arise in supersymmetric models with explicitly
broken R-parity. Using the density and $Y_e$ profiles from Wilson's 
supernova model (see Fig. 1 in ref. \cite{NQRV}), we plot in Fig. 1 
two contours of the conversion probability
in the ($(|\lambda'_{ij1}|^2-|\lambda'_{1j1}|^2),\:
\lambda'_{1j1} \lambda'_{ij1}$) parameter space ($i=2,3; j=1,2,3$). 
Here the {\sl reference} squark mass has been chosen to be 100 GeV.
Should the squark mass be different the plot should be
appropriately re-scaled. 
The solid line is for a conversion probability of $P \approx 0.5$, 
and the dashed one is for $P \approx 0.35$. We see from the figure
that, provided the violation of universality induced by the new
diagonal interactions is sufficiently high that the resonant
conversions take place, i.e. if 
$(|\lambda'_{ij1}|^2-|\lambda'_{1j1}|^2) \gsim 10^{-2}$
one can rule out 
$\lambda'_{1j1} \lambda'_{ij1} \gsim 10^{-6} \div 10^{-4}$. 
Note, that this bound on $\lambda'_{1j1} \lambda'_{ij1}$ is about
three orders of magnitude stronger than the present experimental 
one in \eq{boun}. 

In addition, the region above the neutrinosphere is also
supposed to be the site for the synthesis of heavy elements (with 
mass number $A > 70$) through $r$ processes \cite{Woosley}. 
A necessary condition required for this to occur is $Y_e < 0.5$ 
in the nucleosynthesis region. The value of $Y_e$ is controlled 
by the charged current reactions:
\begin{eqnarray}
\label{nu-n}
\nu_e+n & \rightleftharpoons & p+e^-,\\
\label{nu-p}
\bar\nu_e+p&\rightleftharpoons&n+e^+.
\end{eqnarray}
Roughly speaking, the rates $\Gamma_{\nu_e N}$ 
of the above reactions are proportional to the products
of the $\nu_e$ and $\bar{\nu}_e$ luminosities and average energies, 
\begin{equation}
\label{rates}
\Gamma_{\nu N}\approx \phi_\nu\,\langle \sigma_{\nu N}\rangle
\propto {L_\nu\over\langle E_\nu\rangle}\langle E_\nu^2\rangle
\propto L_\nu\langle E_\nu\rangle \, ,
\end{equation}
where $\phi_\nu$ is the neutrino flux, $\sigma_{\nu N}\propto E_\nu^2$ is
the neutrino absorption cross section, and $\langle\ \rangle$ denotes
the averaging over the neutrino energy distribution. As a result,
 the relevant expression for $Y_e$ turns out to be very simple:
\begin{equation}
\label{fourthye}
Y_e\approx {\Gamma_{\nu_en}\over\Gamma_{\bar\nu_ep}+\Gamma_{\nu_en}}
\approx {1\over 1+\langle E_{\bar\nu_e}\rangle/\langle E_{\nu_e}\rangle}.
\end{equation}
Using the average energies in \eq{average},
we obtain $Y_e\approx 0.41$, in good
agreement with the numerical supernova models.

However, in the presence of neutrino conversions, average energies 
of $\bar\nu_e$ and/or  $\nu_e$ can be affected and consequently the 
value of $Y_e$ can deviate from the predicted one. 

As a result, in the nucleosynthesis region $Y_e$ should
be replaced by
\begin{equation}
\label{34}
Y_e \approx {1\over 1+{\VEV{E_{\bar\nu_e}}_{\rm eff}}/
\VEV{E_{\nu_e}}_{\rm eff}},
\end{equation}
where 
\bea
\VEV{E_{\bar\nu_e}}_{\rm eff} \equiv \VEV{E_{\bar\nu_e}} (1-P) + 
\VEV{E_{\bar\nu_\tau}} P,	\\\nonumber
\VEV{E_{\nu_e}}_{\rm eff} \equiv \VEV{E_{\nu_e}} (1-P) + 
\VEV{E_{\nu_\tau}} P.
\eea
Due to the the {\sl simultaneous} occurrence of resonant
$\nu_e \leftrightarrow \nu_\tau$ and 
$\bar{\nu}_e \leftrightarrow \bar{\nu}_\tau$ conversions, 
there is a trend to equalize the average $\nu_e$ and $\bar\nu_e$
energies, and as a result, to increase $Y_e$ with respect to the 
standard model case with no neutrino or anti-neutrino conversions.

For conversion probabilities of $P\approx0.15$, 0.3, and 0.8, we obtain
$Y_e\approx 0.43,$ 0.45, and 0.49. In Fig. 2, we present the contour lines 
corresponding to these $Y_e$ values. 
The dotted, dashed, and solid
lines in this figure are for $Y_e\approx 0.43$, 0.45, and 0.49, respectively.
If we take $Y_e<0.45$ as a criterion for a successful $r$-process, then  
$\lambda'_{131} \lambda'_{ij1} \gsim 10^{-6} \div 10^{-4}$
is excluded for  $(|\lambda'_{ij1}|^2-|\lambda'_{131}|^2) \gsim 10^{-2}$. 
This excluded region is similar to the previous one obtained
by considering the $\bar\nu_e$ energy spectra from SN1987a, 
because the limits on the conversion probability are about the 
same in both cases. However, we note that if the $r$-process indeed 
occurs in supernovae, then the resulting limits on the effective FCNC 
couplings are much less dependent on the predicted average neutrino energies
than the previous one. This is because the $r$-process argument relies
only on the ratio of the average neutrino energies [cf. \Eq{fourthye}]. 

A remark is in order. The parameter space we have explored in this 
section is complementary to the one relevant for the solar 
neutrino problem \cite{GMP,BPW}. Indeed, in the solar case much larger 
values of the FDNC couplings $(|\lambda'_{331}|^2-|\lambda'_{131}|^2)
\sim 0.4\div 0.6$ are necessary to satisfy the resonance condition in 
the inner solar core where $Y_e \sim 0.7$.  
Certainly, the $\bar{\nu}_e$ energy spectrum consideration 
could be used to exclude, at least partially, the resonant
massless neutrino conversion as a solution to the solar neutrino
problem
\footnote{Note that such solution is already disfavoured, since it 
predicts an energy-independent neutrino suppression, contrary to what 
is indicated by present solar neutrino observations.}, as 
suggested in \cite{AS}. In that case the value of the 
effective FDNC couplings should be much larger in order to
allow the resonant neutrino conversion to take place, i.e.  
($|\lambda'_{331}|^2-|\lambda'_{131}|^2) \geq 0.5$. 
This would correspond to massless resonant neutrino conversion 
very far from the neutrinosphere, unlike the case studied in the
present paper.
On the other hand, no complementary information  can be obtained 
from the r-process nucleosynthesis argument, since this requires 
neutrinos to undergo the resonance just above the neutrinosphere.

\section{Massive Neutrino Conversion in Supernovae}

In models with explicitly broken R parity
neutrino masses are induced radiatively at the one-loop
level due to the exchange of down-type quarks and
squarks  \cite{beyond}. A simple estimate of the corresponding 
diagram shown in Fig. 3, leads to a typical neutrino mass parameter 
$\lambda'^2 m_d^2 /m_{SUSY}$. For reasonable choices 
of $m_{SUSY}$ and $\lambda'$ (see below) one can see that
the resulting neutrino masses could lie in the eV range
for which they could play an important role in neutrino 
propagation in the supernova environment. Moreover, such
mass could account for the hot dark matter in the Universe.
In this section we include the effect of  
non-zero $\delta m^2$ on our previous evolution Hamiltonian
of \eq{hamil}.
Let us assume, for definiteness, that the vacuum mixing angle 
characterizing the 
two-neutrino system is negligible and, moreover, that one of 
the two neutrino species is much heavier than the other.
In our description we will neglect the {\sl non standard} 
FDNC contributions in the Hamiltonian matrix \eq{hamil}, this 
way evading the constraints given in \eq{boun}.
In contrast, the FCNCs generated by the R-parity breaking
interactions provide the required mixing term in the
evolution Hamiltonian, through the matter potential $V_{ex}$. 
In this case the resonant condition reduces to the familiar
one for the MSW effect with vanishing mixing, i.e.
\beq
\frac{\delta m^2}{2E} = \frac{\sqrt2 G_F \rho}{m_p} Y_e
\eeq

A simple numerical check shows that the relevant neutrino mass
scale for which the corresponding resonant neutrino conversions
will occur in the supernova environment includes neutrino mass 
range of few eV, which is precisely the one required in order 
that one of the two neutrino species, \ne or \nt play a role 
as hot dark matter \cite{cobe2}.

The  neutrino wave length is still given by \eq{length} 
where the mixing angle is now given by:
\beq
\label{angle2}
\tan 2\theta_m 
=  \frac{ 2 \xi_{ex}\rho(2-Y_e)}
{\rho Y_e -\delta m^2 m_p /(2\sqrt{2}G_FE)}\, .
\eeq
Therefore, the transition probability is given by 
\bea
\label{LZ2}
  P & = & 1 - 
\exp\Biggl(-\dfrac{\pi^2}{2}\dfrac{\delta r}
{L_{m}^{\rm res}} \Biggr) \nonumber \\
          & \approx & 1 - \exp\left[
-1.6 \times 10^{-2} \times
\left(\dfrac{\delta m^2}{1{\mbox{eV}^2}}\right)
\left(\dfrac{10\mbox{MeV}}{E}\right)
\left(\dfrac{2-Y_e}{Y_e}\right)^2_{res}
            \Biggl(\dfrac{h}{\mbox{cm}} \Biggr)
\xi^2_{ex}
                 \right], \nonumber \\
 \delta r & = & 4 h \xi_{ex}\left(\dfrac{2-Y_e}{Y_e}\right), 
\,\,\,\,\,\,\,\,
h \equiv 
\left| \frac{\mbox {d}\ln (\rho Y_e)}{\mbox{d}r}\right|^{-1}_{\rm res}, 
\end{eqnarray}

This way we will constrain the $(\delta m^2, 
\lambda'_{1j1}\lambda'_{ij1})$ parameter space irrespective of 
any universality violation.

Let us note that for a given sign
\footnote{Here we set $\delta m^2 > 0$ for $m_{\nu_x} > m_{\nu_e}$.}
 of $\delta m^2 $  only one kind of resonant conversion, 
either $\nu_e\leftrightarrow\nu_x$ (for $\delta m^2>0$), or 
$\bar{\nu}_e\leftrightarrow\bar{\nu}_x$ (for $\delta m^2<0$), can occur.
Therefore to discuss $\bar{\nu}_e$ energy spectra distortion 
from SN1987a we have to assume $\delta m^2<0$. 
The upper bound on $\bar{\nu}_e$ mass from $\beta$ decay experiment, 
$m_{\nu_e} < 4.35$ eV (95\% C.L.) \cite{beta} cut off our relevant
$\delta m^2$ range in Fig. 3. One sees from this figure that for 
$\delta m^2 \lsim 1 \div 20$eV$^2$ 
the FCNC couplings are restricted to be $\lsim 10^{-3}$
irrespective of any lepton non-universality. From this point of view 
the limits derived 
in this section are of more general validity than those of section 3. 
For this mass hierarchy the resonant neutrino conversion would not 
conflict with the nucleosynthesis process for any choice of
parameters, and therefore no constraint can be obtained. 

On the other hand, for $\delta m^2>0$ one expects that 
$\nu_e \leftrightarrow \nu_{x}$ transitions will occur 
and they  can affect the nucleosynthesis process. 
In contrast, in this case the $\bar{\nu}_e$ spectra 
would be unaffected. In Fig. 4 we plot the iso-contours 
for different values of the electron abundance $Y_e$. 
One can see  that in the interesting range 
$\delta m^2 \sim 1 \div 20 $eV$^2$,  
favoured by the hot plus cold dark matter scenario \cite{cobe2}, 
we can  rule out the FCNC couplings $\lambda'_{1j1} \lambda'_{ij1}$  
at the level of few $10^{-3}$. 

\section{Resonant Massless Neutrino Conversion and Supernova Shock Re-heating}

We would like to briefly address an interesting open problem 
related with the energetics of supernova explosion. 
It is now generally accepted that the prompt shock stalls at a 
radius $\sim 100$ kilometres, due to photo-dissociation, 
neutrino losses, and accretion \cite{burrows}. The main aspect 
of a supernova explosion is the transfer of energy from the core 
to the mantle.  The mantle is less bound than the core, whose 
binding energy can grow during the delay to 
explosion.  The core is the protoneutron star that will
evolve due to neutrino cooling and deleptonization over 
many seconds. Bethe \& Wilson \cite{bw85} showed how neutrino
heating of the accreted material near the shock could lead to an 
explosion.  It seems compelling that neutrinos mediate this energy 
transfer and are the agents of explosion \cite{burrows}.

If neutrinos have only standard model interactions the energy they
carry seems insufficient to re-energyse the shock material.
It has been argued that the occurrence of $\nu_e \ra \nu_{\mu,\tau}$ 
MSW neutrino conversions behind the shock would increase the energy 
deposited by $\nu$`s. This is due to the fact that the average energy 
of $ \nu_{\mu,\tau}$ is about twice larger than that of $\nu_{e}$. 
The capture processes in \eq{nu-n} and \eq{nu-p} are mostly responsible 
for the energy deposit. 

Our scenario is rather distinct from the MSW effect. 
Unlike in the MSW case, the simultaneous $\nu_e \ra \nu_{\mu,\tau}$ and 
$\bar\nu_e \ra \bar\nu_{\mu,\tau}$ conversions can power 
both reactions \eq{nu-n} and \eq{nu-p} and as a result the effect 
may be larger than  for the standard MSW or resonant spin-flavour 
precession \cite{FMMW,ALPS}.
 
We adopt the argument given by Fuller {\it et al.} in \cite{FMMW} 
for providing the total heating rate by $\nu_e$ and $\bar{\nu}_e$. 
Qualitatively, the heating rate $\dot{\epsilon}$ is just the product 
$\vev{E} \Gamma_{\nu N} Y_N$ (see \eq{rates}), namely 
\beq
\dot{\epsilon} \approx L_{\nu} \biggl(Y_n \langle E_{\nu_e}\rangle^2  + 
Y_p \langle E_{\bar{\nu}_e}\rangle^2\biggr) 
\eeq
In the presence of complete resonant conversions $\nu_{\mu}\ra \nu_e$ and 
$\bar{\nu}_{\mu}\ra \bar{\nu}_e$ the rate can be increased by the amount
\beq
\label{ratio}
\frac{\dot{\epsilon}'}{\dot{\epsilon}} \approx 
\frac{Y_n \langle E_{\nu_\tau} \rangle^2  + 
Y_p \langle E_{\bar{\nu}_\tau}\rangle^2} 
{Y_n \langle E_{\nu_e} \rangle^2  + 
Y_p \langle E_{\bar{\nu}_e}\rangle^2} = 
\biggl(\frac{\langle E_{\nu_\tau} \rangle}
{\langle E_{\nu_e} \rangle}\biggr)^2 \sim 2\, ,
\eeq
where it is assumed  $\langle E_{\nu_\tau} \rangle = 
\langle E_{\bar{\nu}_\tau} \rangle \sim 21$ MeV 
and $\langle E_{\nu_e} \rangle = 
\langle E_{\bar{\nu}_e}\rangle \sim 15$ as typical average energies for 
the earlier epoch after the bounce $t \gsim 0.1$ s. 
At this epoch, the $Y_e$ value is somewhat larger than 
that characteristic of the later epoch discussed above $Y_e 
\sim 10^{-2}$. 
However, the present experimental bounds on 
$\lambda'_{1j1}, \lambda'_{3j1}$ allow $\xi' \gsim 0.1$, 
needed in order to have resonant neutrino conversions (see \eq{rc})
at $t\gsim 0.1$ s if $Y_e \sim 0.15$ at neutrino sphere.

We can notice that in the usual $\nu_e \leftrightarrow \nu_x$ 
MSW conversion 
\footnote{
Our estimates of the heating rates are somewhat qualitative 
but they are sufficient for our discussion.} the gain in reheating 
rate with respect to that of the standard model is 
\cite{FMMW} ${\dot{\epsilon}'}/{\dot{\epsilon}} \approx 5/3$  
whereas in the resonant spin-flavour precession scenario 
\cite{ALPS} ${\dot{\epsilon}'}/{\dot{\epsilon}} \approx 4/3$.

Clearly, for the massive neutrino  case  we can also expect 
analogous effects. Actually the scenario, depending on 
the sign of $\delta m^2$ looks like the usual MSW picture.
    
\section{Conclusions}

Supersymmetry with explicitly broken $R$ parity breaking 
provides a variety of novel possibilities for  neutrino 
propagation properties in the presence of matter, even
when they are strictly massless. 
The supernova matter background 
seems to be one where most likely resonant conversions of  
massless neutrinos can play an important role.

We have re-examined the resonant massless-neutrino conversion 
in a supernova medium in the presence of flavour changing neutral 
current (FCNC) couplings present in explicit $R$ parity violating 
supersymmetric models.
We have shown how the observed $\bar\nu_e$ energy spectra 
from SN1987a and the supernova $r$-process nucleosynthesis  argument 
may provide very stringent constraints on such new FCNC interactions.
Typically they are much more stringent than previously obtained at the
laboratory. From this point of view the SN1987a event provides
 a strong sensitivity in restricting  neutrino properties
in supersymmetric models with $R$ parity violation.
Our results here are summarized in Figs. 1 and 2.

We have also generalysed the description of  MSW massive-neutrino 
conversions in supernovae so as to account for the  presence of 
explicit $R$-parity-violating  FCNCs and determined the corresponding 
restrictions in the limit of vanishing vacuum mixing. Our results are 
summarized in Figs. 3 and 4. The relevant
neutrino mass scale could play an important role in connection with 
hot dark matter. While these constraints we derive on $R$ parity 
violating interactions are weaker than the ones obtained in the
massless limit they are still stronger than those available
from laboratory experiments. More importantly, they 
are of wider validity than those obtained in the
massless limit.

Last but nor least,  our discussion of massless-neutrino 
conversions in supernovae should highlight the interest in 
improving the present laboratory limits on universality 
violation and flavour changing R-parity breaking interactions.


\centerline{\bf Acknowledgement}
\noindent
We thank Alexei Smirnov for fruitful discussions.
This work was supported by DGICYT under Grant  PB92-0084, 
by the Human Capital and Mobility Program under Grant  
ERBCHBI CT-941592 (A. R.), and by a DGICYT postdoctoral 
fellowship (H. N.).

\noindent

\bef
\vglue -1.5cm
\centerline{\protect\hbox{
\psfig{file=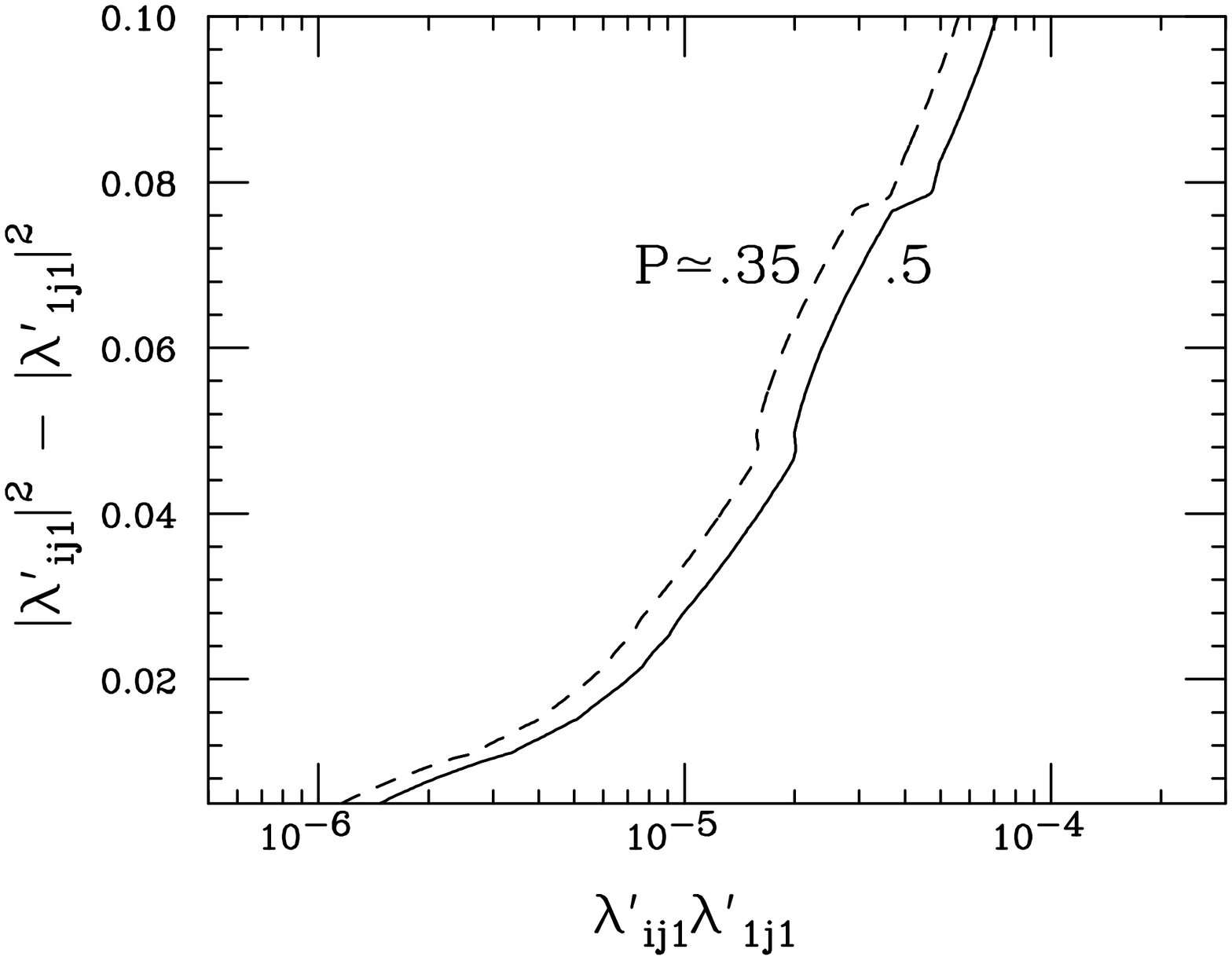,height=10.0cm,width=13.0cm,angle=90}}}
\vglue -1.0cm
\caption{
Constraints on the $R$-parity violating couplings from the observed SN1987a 
$\bar\nu_e$ energy spectra.  Here, $i=2,3; j=1,2,3$. The  dashed (solid) 
lines correspond to an allowed conversion probability of $P = 0.35$ (0.5). 
The region to the right of these lines are excluded by the requirement 
$P<0.35$ (0.5), as indicated by the SN1987a data.}
\vglue 1.0cm
\centerline{\protect\hbox{
\psfig{file=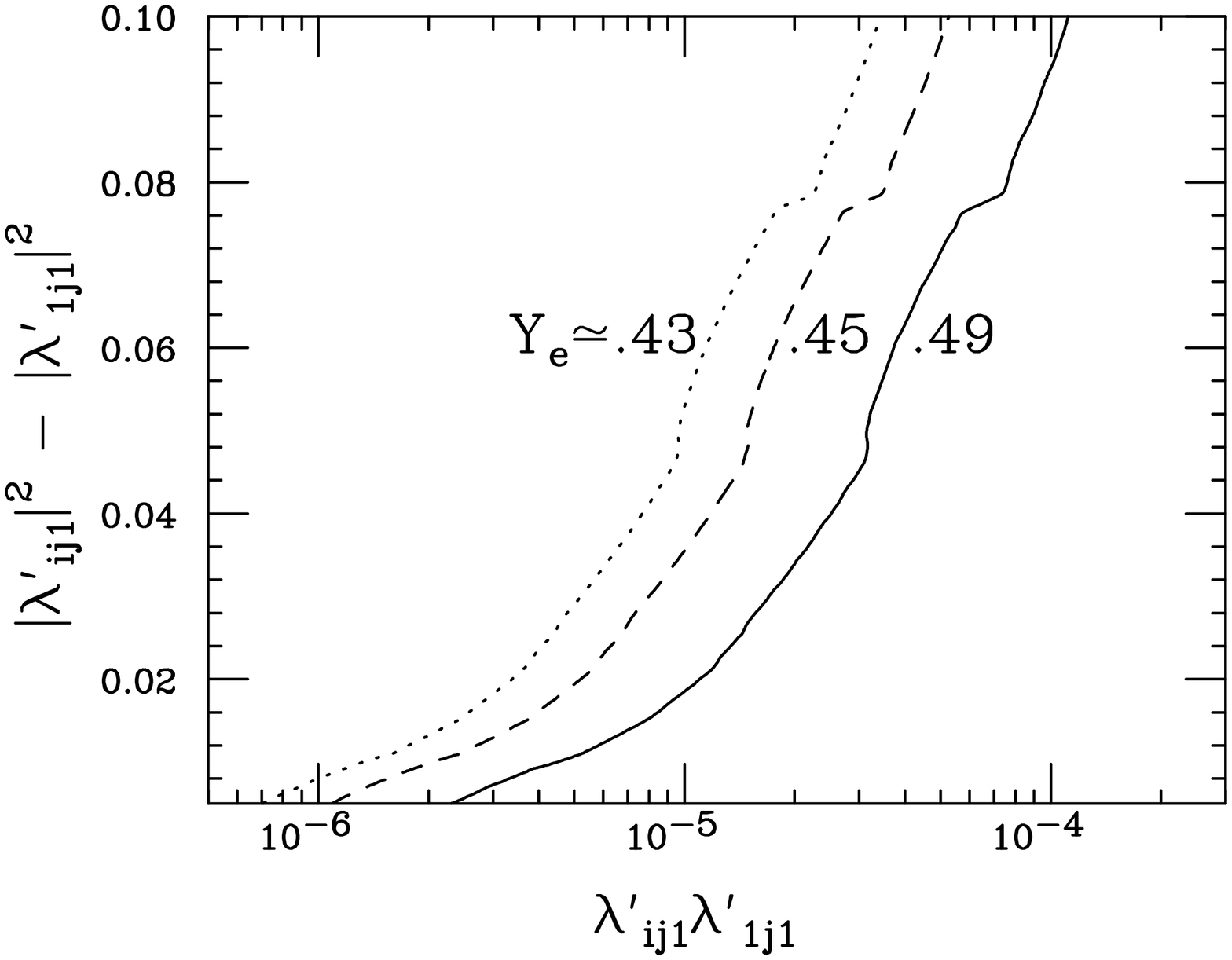,height=10.0cm,width=13.0cm,angle=90}}}
\vglue -1.0cm
\caption{Constraints on the $R$-parity violating couplings
from the supernova $r$-process nucleosynthesis. The region to the
right of the dotted, dashed and solid lines are exclued for the 
required values of $Y_e<0.43$, 0.45, and 0.49, respectively,
in the $r$-process. }

\eef

\bef
\vglue -1.5cm
\centerline{\protect\hbox{
\psfig{file=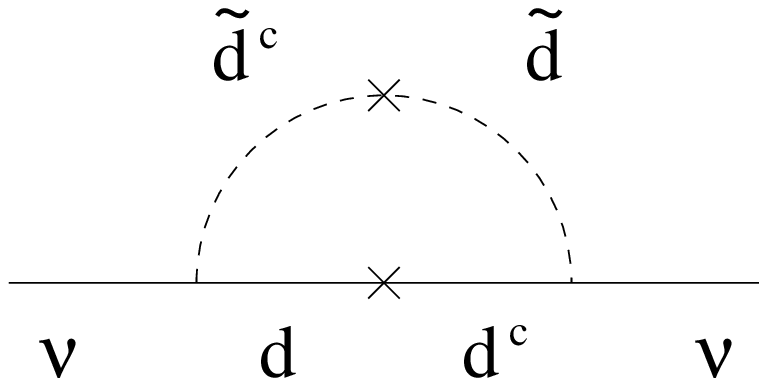,width=13.0cm}}}
\vglue -1.0cm
\caption{
Typical diagram generating neutrino mass in a supersymmetric 
model with explicitly broken R-parity.}
\eef

\bef
\vglue -1.5cm
\centerline{\protect\hbox{
\psfig{file=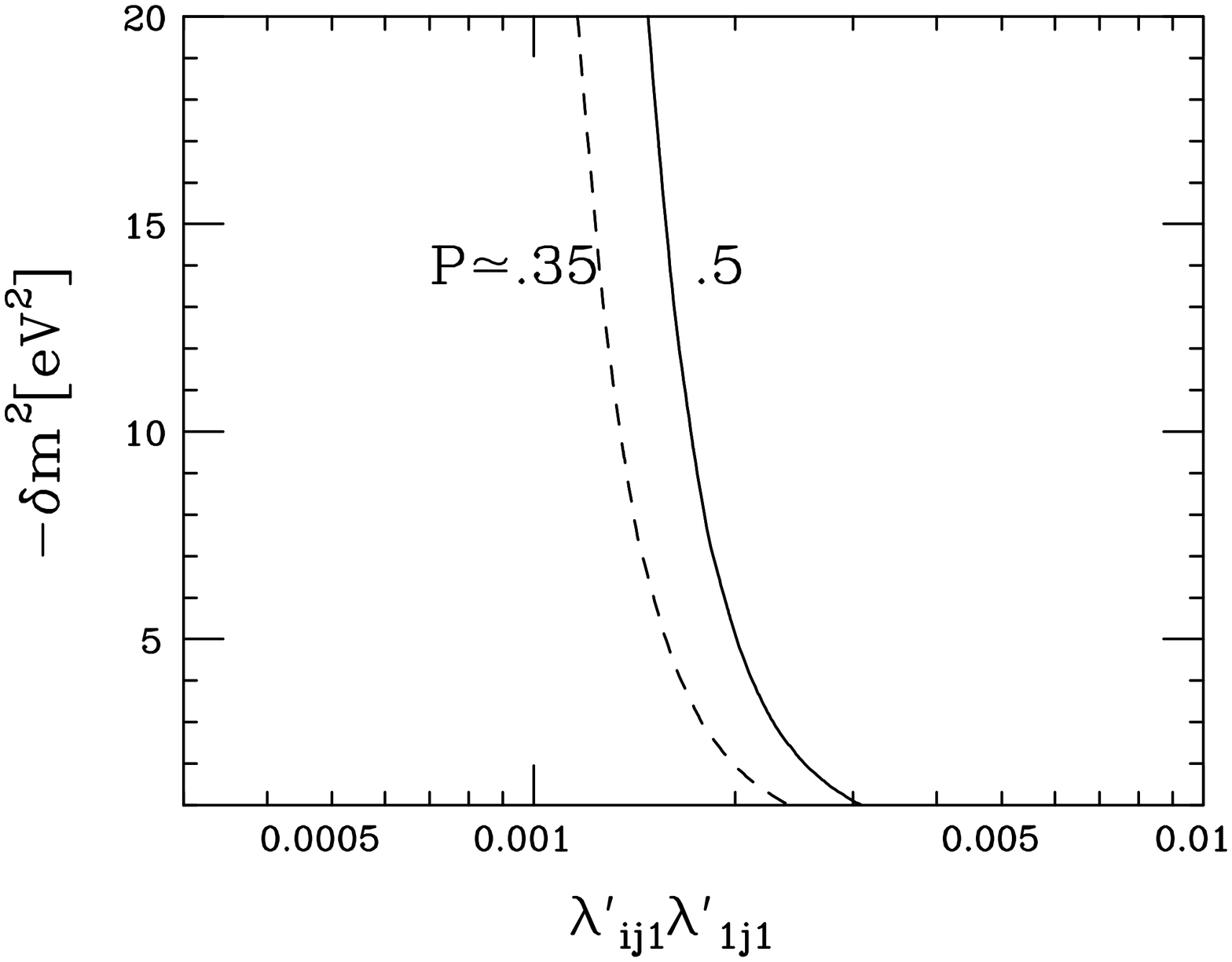,height=10.0cm,width=13.0cm,angle=90}}}
\vglue -1.0cm
\caption{
SN1987a $\bar\nu_e$ energy spectra constraints on the 
FCNC $R$-parity violating couplings for as a function of 
$\delta m^2$ and for negligible vacuum neutrino mixing. 
The region to the right of the dashed (solid) lines are 
excluded by the data for an allowed conversion probability 
of $P<0.35$ (0.5) irrespective of any laboratory restriction 
on R-parity-violating intereactions.}
\vglue 1.0cm
\centerline{\protect\hbox{
\psfig{file=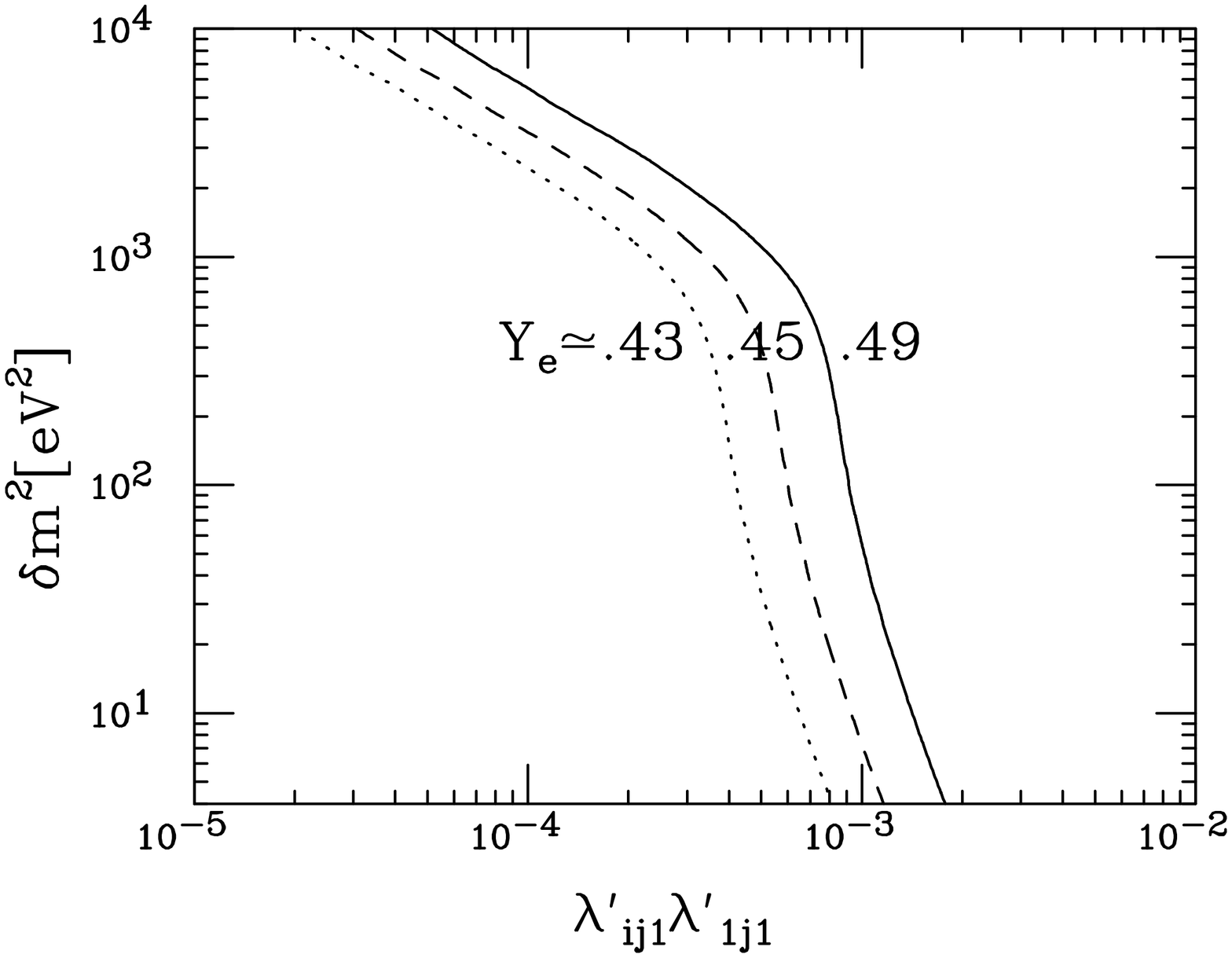,height=10.0cm,width=13.0cm,angle=90}}}
\vglue -1.0cm
\caption{Constraints on the $R$-parity violating couplings
from the supernova $r$-process nucleosynthesis. The region to the
right of the dotted, dashed and solid lines are exclued for the 
required values of $Y_e<0.43$, 0.45, and 0.49, respectively,
in the $r$-process. }

\eef


\begin{thebibliography}{99}

\bibitem{W} 
L. Wolfenstein,   \pr{D17}{2369}{78}. 

\bibitem{valle}
J. W. F. Valle,  \pl{B199}{432}{87}. 

\bibitem{ER} 
E. Roulet, \pr{D44}{R2935}{91}. 

\bibitem{GMP}
M. M. Guzzo, A. Masiero and S.T. Petcov, \pl{B260}{154}{91}. 

\bibitem{MS}
S. P. Mikheyev and A. Yu. Smirnov, \sjnp{42}{913}{85}. 

\bibitem{AS}
A. Yu. Smirnov, in {\it Proceedings of the 4th Int. Symposium on 
Neutrino Telescopes}, Venice, Italy, ed. by M. Baldo Ceolin (University of 
Padova, 1992), p.63.

\bibitem{NQRV}
H. Nunokawa, Y.-Z. Qian, A. Rossi and J. W. F. Valle, 
Valencia preprint FTUV/96-25, IFIC/96-29; hep-ph/9605301.

\bibitem{BPW}
V. Barger, R. J. N. Phillips and K. Whisnant\pr{D44}{1629}{91};
S. Degl' Innocenti and B. Ricci, \mpl{A8}{471}{93}.

\bibitem{beyond}
For a recent review see J. W. F. Valle, in
{\sl Physics Beyond the Standard Model}, lectures given at the 
{\sl VIII Jorge Andre Swieca Summer School} (Rio de Janeiro, February 1995) 
and at {\sl V Taller Latinoamericano de Fenomenologia de las Interacciones
Fundamentales} (Puebla, Mexico, October 1995); hep-ph/9603307 

\bibitem{SW} 
S. Weinberg, \pr{D26}{287}{82}. 

\bibitem{rparity} 
C. Aulak and N. R. Mohapatra, \pl{B119}{136}{83};\\
F. Zwirner, \pl{B132}{103}{83};\\
L. J. Hall and M. Suzuki, \np{B231}{419}{84}; \\
J. Ellis {\it et al.}, \pl{B150}{142}{85}; \\
G. G. Ross and J. W. F. Valle, \pl{B151}{375}{85};\\
R. Barbieri and A. Masiero, \pl{B267}{679}{86}. 

\bibitem{QFMMW}
Y.-Z. Qian {\it et al.,} \prl{71}{1965}{93}.

\bibitem{old}
P. Reinartz and L. Stodolsky,  \Zp{C27}{507}{85};\\
L. Wolfenstein,  \pl{B194}{197}{87}.

\bibitem{SSB}
A. Yu. Smirnov,  D. N. Spergel and J. N. Bahcall, 
\pr{D49}{1389}{94}.

\bibitem{JNR} 
B. Jegerlehner, F. Neubig and G. Raffelt, preprint MPI-PTh 95-120, 
astro-ph/9601111.

\bibitem{mssm}
See for a review, e.g., 
 H. Haber and  G. Kane, \prep{117}{85}{75};
 H. P. Nilles, \prep{110}{84}{1};
R. Barbieri, \r.n.c.{11}{1}{88}.

\bibitem{BGH}
V. Barger, G. F. Giudice and T. Han, \pr{D40}{2987}{89}; \\
C. E. Carlson, P. Roy and  M. Sher, \pl{B357}{99}{95}; \\
D. Choudhury and P. Roy, preprint MPI-PTh/96-20, TIFR/TH/96-12, 
hep-ph/9603363.

\bibitem{VS}
A. Yu. Smirnov and F. Vissani, preprint IC/96/16; hep-ph/9601387 

\bibitem{GB}
G. Bhattacharyya, preprint  hep-ph/9511447, to appear in the 
{\it Proceedings of the HEP Conference}, Brussels, 1995.

\bibitem{KP} 
M. Hirsch, H.V.Kapodor-Kleigrothaus and S.G. Kovalenko
\pl{B372}{181}{96} 

\bibitem{Landau} 
L. Landau,  Phys. Z. Sowjetunion {\bf 2}, 46 (1932);\\
C. Zener,  Proc. R. Soc. London {\bf A137}, 696 (1932).

\bibitem{HPD}
W. C. Haxton, 
 Phys. Rev. Lett. {\bf 57}, 1271 (1986); \\
S. J. Parke, 
Phys. Rev. Lett. {\bf 57}, 1275 (1986); \\
A. Dar {et al.},  Phys. Rev. {\bf D35}, 3607 (1987).  

\bibitem{KA}
K. Hirata {\it et. al.}, \prl{58}{1490}{87}.

\bibitem{IMB}
R. Bionta {\it et. al.}, \prl{58}{1494}{87}.

\bibitem{Woosley}
S. E. Woosley and E. Baron,  Astrophys. J. {\bf 391}, 228 (1992); \\
S. E. Woosley,  Astron. Astrophys. Suppl. Ser. {\bf 97}, 205 (1993);\\
S. E. Woosley and R. D. Hoffman,  Astrophys. J. {\bf 395}, 202 (1992);\\
B. S. Meyer {\it et al.,} Astrophys. J.  {\bf 399}, 656 (1992); \\
S. E. Woosley {\it et al.}, Astrophys. J. {\bf 433}, 229 (1994).

\bibitem{cobe2}
For reviews, see R. Rowan-Robinson, in {\sl Cosmological Dark Matter}, 
(World Scientific, 1994), ed. J. W. F. Valle and A. Perez, p. 7;
C. S. Frenk, {\sl op. cit.}, p. 65; 
J. Primack {\sl op. cit.}, p. 81.

\bibitem{beta}
A. I. Belesev {\it et al.}, \pl{B350}{263}{95}.

\bibitem{burrows}
For a recent overview and earlier references, see A. Burrows, 
astro-ph/9606035.

\bibitem{bw85}
H. A. Bethe, J. R. Wilson, \apj{307}{178}{86}

\bibitem{FMMW}
G. M. Fuller {\it et al.}, Astrophys. J. {\bf 389}, 517 (1992).  

\bibitem{ALPS}
E. Kh. Akhmedov, A. Lanza, S.T. Petcov and D.W. Sciama, 
SISSA-40-96-A-EP; TUM-HEP-239-96;  hep-ph/9603443 

\end{thebibliography}
\end{document}